\documentclass{svmult}

\usepackage{makeidx}         
\usepackage{graphicx}        
\usepackage{multicol}        
\usepackage[bottom]{footmisc}

\usepackage{amsmath,amssymb}

\makeindex             

\renewcommand{\>}{\rangle}
\newcommand{\uk}{|\kern-3.3pt\uparrow\>}
\newcommand{\dk}{|\kern-3.3pt\downarrow\>}

\begin{document}

\title*{Occupation time for classical and quantum walks}

\author{F. A. Gr\"unbaum\inst{1}\and L. Vel\'{a}zquez\inst{2}\thanks{
    The work of the second author has been supported in part by
  the research project MTM2017-89941-P from Ministerio de
  Econom\'{\i}a, Industria y Competitividad of Spain and the European
  Regional Development Fund (ERDF), by project UAL18-FQM-B025-A
  (UAL/CECEU/FEDER) and by project E48\_20R of Diputaci\'on General de
  Arag\'on (Spain) and the ERDF 2014-2020 ``Construyendo Europa desde
  Arag\'on.''}
  \and
  J. Wilkening\inst{1}\thanks{The work of the third author was supported in part by
  the US National Science Foundation under award number DMS-1716560 and
  by the US Department of Energy, Office of Science, Applied Scientific
  Computing Research, under award number DE-AC02-05CH11231.}
}

\institute{Department of Mathematics, University of California, Berkeley, CA, 94720
\texttt{grunbaum@math.berkeley.edu},
\texttt{wilkening@berkeley.edu} \\
\and Departamento de 
Matem\'{a}tica Aplicada, Universidad de Zaragoza, Zaragoza, Spain
\texttt{velazque@unizar.es}}
%
%

\maketitle

\begin{abstract}
This is a personal tribute to Lance Littlejohn on the occasion of his
70\textit{th} birthday.  It is meant as a present to him for many years of
friendship.  It is not written in the ``Satz-Beweis'' style of Edmund
Landau or even in the format of a standard mathematics paper.  It is
rather an invitation to a fairly new, largely unexplored, topic in the hope
that Lance will read it some afternoon and enjoy it.  If he cares
about complete proofs he will have to wait a bit longer; we almost
have them but not in time for this volume.  We hope that the figures
will convince him and other readers that the phenomena displayed here
are interesting enough.
\end{abstract}

\section{Introduction}

The origin of our story goes back to a remarkable observation by Paul
Levy \cite{Le}, that one dimensional Brownian motion behaves in rather
unexpected ways. To state his result in everyday terms let us switch
to a long night at the casino where you play repeatedly with a fair
coin: if you get heads you win one dollar, if you get tails you lose
one dollar. You play once every ten seconds and you keep track of your
``fortune" as a function of time: you are happy if your fortune at
that time is positive, i.e. you have won more times than you have
lost, otherwise you are unhappy. This game has to be played for a
fixed and large number of tosses. What proportion of the total time
you play do you expect to be happy? It is not unreasonable to guess
that the answer should be about $1/2$.

\medskip

However, and this is P.~Levy's observation, a ``wild night" is more
likely than a ``normal night". Take---for instance---two windows of
total length $1/10$: then the probability that you are happy somewhere
between $45$ and $55$ percent of the time is smaller than the
probability that you are happy either more than $95$ or less than $5$
percent of the time. This discussion can be found in many standard
probability books, \cite{feller,McK,S1}. The distribution of the
random variable in question is explicitly known---and is given by an
arcsine law---and as evidence of its interest we observe that one of
the first applicaction of the Feynman-Kac formula was a rederivation
of this result of P.~Levy by M.~Kac, see \cite{Ka}. The careful
formulation of the discrete version with Brownian motion replaced by a
coin was done by K.~Chung and W.~Feller, see \cite{CF}.

\medskip

Just to put things in perspective, consider two-dimensional Brownian
motion starting at the origin, run it for time one and inquire about
the time spent in the positive quadrant. The analytical form of the
result is not known. See \cite{GM,ES} and references there.

\medskip

The rest of this paper is devoted to a discussion of what happens when
one replaces a classical random walk on the integers, such as the
ordinary coin, with a quantum walk. This is not the place to give a
full account of this notion. We just mention the paper by Y. Aharonov,
L. Davidovich and N. Zagury, \cite{Ah}, where this was first
introduced; a very good and early survey paper by Julia Kempe,
\cite{K}; and the first paper where this was connected with orthogonal
polynomials on the unit circle, see \cite{CGMV}. It is also useful to
look at \cite{Ko}.

\medskip

We will study a few quantum walks whose state space is a Hilbert space
spanned by an orthonormal set of states $|i\>\otimes\uk$,
$|i\>\otimes\dk$ with a definite ``site" $i$ along the integers and a
definite value of an extra degree of freedom---the ``spin"---represented
by an up or down arrow, which may be viewed as the quantum
counterpart of the two sides of a coin. At each time step, the
evolution is driven by some ``transition rules"
\[
\begin{aligned}
  |i\>\otimes\uk \longrightarrow \begin{cases}
    |i+1\>\otimes\uk & \text{with (complex) probability amplitude} \hskip 5pt c_{11}^i
    \\
    |i-1\>\otimes\dk & \text{with (complex) probability amplitude} \hskip 5pt c_{21}^i
  \end{cases}
  \\
  |i\>\otimes\dk \longrightarrow \begin{cases}
    |i+1\>\otimes\uk & \text{with (complex) probability amplitude} \hskip 5pt c_{12}^i
    \\
    |i-1\>\otimes\dk & \text{with (complex) probability amplitude} \hskip 5pt c_{22}^i
  \end{cases}
\end{aligned}
\]
where, for each $i\in\mathbb{Z}$,
\begin{equation} 
  C_i = \begin{pmatrix} c_{11}^i & c_{12}^i \\ c_{21}^i & c_{22}^i \end{pmatrix}
\end{equation}
is an arbitrary unitary matrix which we will call the $i^{th}$
coin. This defines in the Hilbert state space a unitary operator $U$
governing the one-step evolution, so that a walker originally in the
state $\psi$ is in the state $U^n\psi$ after $n$ steps.

\medskip

This kind of ``coined" walk is the simplest of all models. In
\cite{CGMV} this is extended by allowing a few more ``local
transitions" so as to obtain the doubly infinite version of a general
CMV matrix---see \cite{FIVE,CGMV}---as the unitary matrix that gives the
discrete time evolution in the basis $|i\>\otimes\uk$,
$|i\>\otimes\dk$.
The most general doubly infinite CMV matrix has the form
\[
\begin{array}{cccc|cccc}
  \dots & \kern3pt\dots & \kern3pt\dots & \kern3pt\dots
  \kern3pt & \kern3pt
  \dots & \kern3pt\dots & \kern3pt\dots & \kern3pt\dots
  \\
  \dots & \kern3pt \rho_{-3}\overline{\alpha}_{-2} & \kern3pt 
  -\alpha_{-3}\overline{\alpha}_{-2} & \kern3pt 
  \rho_{-2}\overline{\alpha}_{-1}
  \kern2pt & \kern2pt
  \rho_{-2}\rho_{-1} & \kern3pt 0 & \kern3pt 0 & \kern3pt \dots
  \\
  \dots & \kern3pt \rho_{-3}\rho_{-2} & \kern3pt 
  -\alpha_{-3}\rho_{-2} & \kern3pt 
  -\alpha_{-2}\overline{\alpha}_{-1}\;
  \kern2pt & \kern2pt
  -\alpha_{-2}\rho_{-1} & \kern3pt 0 & \kern3pt 0 & \kern3pt \dots
  \\ \hline
  \dots & \kern3pt 0 & \kern3pt 0 & \kern3pt \rho_{-1}\overline{\alpha}_0
  \kern2pt & \kern2pt
  -\alpha_{-1}\overline{\alpha}_0 & \kern3pt \rho_0\overline{\alpha}_1 
  & \kern3pt \rho_0\rho_1 & \kern3pt \dots
  \\
  \dots & \kern3pt 0 & \kern3pt 0 & \kern3pt \rho_{-1}\rho_0 & 
  -\alpha_{-1}\rho_0
  \kern2pt & \kern2pt
  \kern3pt -\alpha_0\overline{\alpha}_1 & \kern3pt -\alpha_0\rho_1 & 
  \kern3pt \dots
  \\
  \dots & \kern3pt \dots & \kern3pt \dots & \kern3pt \dots
  \kern2pt & \kern2pt
  \dots & \kern3pt \dots & \kern3pt\dots & \kern3pt\dots
\end{array}
\]
where $\rho_j=\sqrt{1-|\alpha_j|^2}$ and
$(\alpha_j)_{j=-\infty}^\infty$ is a sequence of complex numbers such
that $|\alpha_j|<1$. The coefficients $\alpha_j$ are known as the
Verblunsky coefficients. A quantum walk whose unitary step in the
basis $|i\>\otimes\uk$, $|i\>\otimes\dk$ is a doubly infinite CMV will
be called a CMV walk.

\medskip

CMV walks are the quantum analogue of the classical birth--death
processes. The Jacobi matrices underlying the latter provide the
canonical representations of self-adjoint operators, while a similar
role in the unitary case is played by the CMV matrices. Also, the
fruitful connection between birth-death processes and orthogonal
polynomials on the real line via Jacobi matrices becomes a similarly
useful connection between CMV walks and the theory of orthogonal
polynomials on the unit circle, the breeding ground where CMV matrices
were born.

\medskip 

CMV walks not only constitute simple quantum walk models, but they are
universal models for quantum walks, since any unitary operator---as the
unitary step governing the evolution of a quantum walk---has a CMV
representation. This paves the way to the use of the CMV
representation as a resource for the analysis of general quantum
walks, a technique nowadays known as the ``CGMV method"
\cite{CGMV2,KoSe,KoSe2}.

\medskip 

Up to a change of phases in the basis, coined walks correspond to CMV
walks whose Verblunsky coefficients with odd index are zero
\cite{CGMV,CGMV2}. The corresponding coins are given by
\begin{equation} \label{eq:coin}
 C_i = 
 \begin{pmatrix} 
 \rho_{2i} & -\alpha_{2i} \\ \overline\alpha_{2i} & \;\rho_{2i} 
 \end{pmatrix}. 
\end{equation}

\medskip 

The examples that we will consider are mostly coined walks: the
Hadamard walk of \cite{Ah,K} which illustrates the behaviour for an
unbiased constant coin, several instances of biased constant coins,
and even a case of a site dependent coin as the the one indicated
above. The last example that we consider here, the Riesz walk, not
only requires going beyond coined walks to consider a CMV walk, but
also involves a non-trivial site dependence for the corresponding
Verblunsky coefficients.

\section{A look at the classical discrete case}

In the discrete case we say that the partial sum $S_k$ (i.e. your
  fortune at time $k$) is positive if $S_k > 0$ or in case that
$S_k=0$ one has $S_{k-1} > 0$. We put $S_0=0$.
If $N_n$ denotes the number of
``positive'' terms among $S_1,S_2,\dots,S_n$ the result of
Chung--Feller \cite{CF} is that the probability that $N_{2n}=2r$ is
\[
  P(N_{2n} = 2r) \; = \; \frac1{2^{2n}}\binom{2r}{r}  \binom{2n-2r}{n-r}\,.
\]
Using the notation $u_{2k}=\frac1{2^{2k}}\binom{2k}{k}$ we have 
\[
  P(N_{2n} = 2r) \; = \; u_{2r} u_{2n-2r}\,.
\]
In \cite{G} one finds an adaptation of the Feynman-Kac method to
reprove the results of Chung and Feller in the discrete case. This
leads to supplement the results in \cite{CF} with

\begin{equation}
  P(N_{2n+1}=2r)\; = \; u_{2r} u_{2n+2-2r} \frac{n-r+1}{n+1}, \qquad r=0,1,2,\dots,n
\end{equation}
and 
\begin{equation}
  P(N_{2n+1}=2r-1) \; = \; u_{2r} u_{2n+2-2r} \frac{r}{n+1}, \qquad r=1,2,\dots,n+1\,.
\end{equation}

\bigskip

For those who have played with Legendre polynomials, such as Lance, may
enjoy the fact that they are connected with this coin-tossing problem.
If we denote the Legendre polynomials by $P_n$ (with $P_n(1)=1$) we
get
\[
 \sum_{k=0}^{n} u_{2k} u_{2n-2k} q^{2k} =
 P_n\left( \frac{q+q^{-1}}{2} \right) q^n.
\]
This connection of the Legendre polynomials and the coin-tossing game is  
essentially in the literature; see \cite{R}, formula(10), page 248.

\bigskip

\begin{figure}[h]
  \centering
  \includegraphics[width=\linewidth]{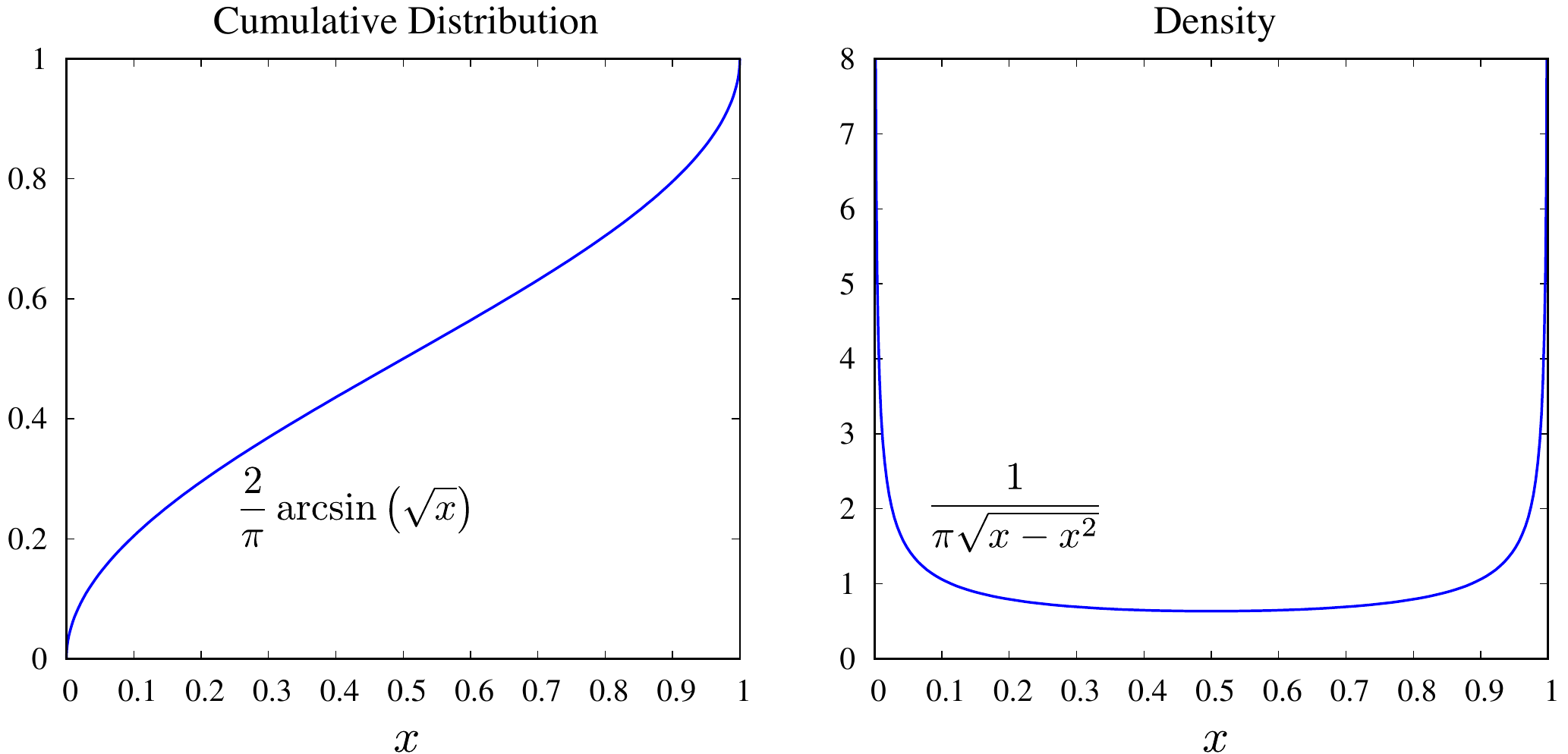}
  \caption{
    Cumulative distribution and
    density for the occupation time of the positive half line under
    Brownian motion.
  }
  \label{fig:clas}
\end{figure}

In the classical case, either the case of Brownian motion or of a fair
coin (in the limit of an infinite number of tosses), the cumulative
distribution and density functions are well-known; see
Figure~\ref{fig:clas}.  The numerical simulations reported in
\cite{GM} agree almost perfectly with this analytical result. Most
of the numerical work in \cite{GM} deals with the two dimensional case
and the occupation time in a wedge, in which case no analytical
results are known. The way that Brownian motion is simulated in
\cite{GM} goes back to P.~Levy himself in terms of a random ``Fourier
series" using Haar functions to approximate white noise.

\section{Occupation times for quantum walks}

When discussing occupation times for quantum walks one should decide
when the walker will be ``observed", since this alters the natural
evolution due to Born's rule about quantum measurements. Here we
consider a ``monitoring" approach to occupation times similar to that
one proposed in \cite{GVWW} and further considered in \cite{bourg} to
study recurrence in quantum walks. One should also look at \cite{KB}.

\medskip 

This gives a natural way to look at the times when a coined or CMV
quantum walk is on each half of the state space. The positive half
consists of the span of the states
\begin{equation} 
  |0\>\otimes\dk\;,\;|1\>\otimes\uk\;,\;|1\>\otimes\dk\;,\;|2\>\otimes\uk\;,\; |2\>\otimes\dk\;,\;\dots
\end{equation}
while the negative half consists of the span of the states
\begin{equation} 
  \dots\;,\;|-2\>\otimes\uk\;,\;|-2\>\otimes\dk\;,\;|-1\>\otimes\uk\;,\;|-1\>\otimes\dk\;,\; |0\>\otimes\uk
\end{equation}
We call these the positive and the negative subspaces of our Hilbert
state space respectively.  In accordance with the setup in
\cite{CGMV,bourg}, the orthogonal projection $P$ on the positive
subspace plays a fundamental role in our results.

\medskip

The monitoring approach assumes that a measurement is performed after
each step to decide whether the walker is on the positive subspace or
not. From a mathematical point of view, this results in the
application of one of the projections, $P$ or $Q=I-P$, after the
unitary step $U$ driving the evolution. While $P$ conditions on the
event ``the walker has been found on the positive subspace", the
complementary projection $Q$ conditions on the event ``the walker has
not been found on the positive subspace". This gives the probability
of any occupation distribution according to Born's rule.

\medskip

For example, if the walker starts at the state $\psi$, $\|\psi\|=1$,
and runs for 6 steps, the quantity
\[
 \|PUPUQUPUQUPU\psi\|^2
\] 
gives the probability to find the walker on the positive subspace at
times 1, 3, 5 and 6, but not at times 2 and 4. As a consequence, the
probability of finding the walker, for instance, 2 times on the
positive subspace, while running for a total number of 4 steps, is
given by the sum
\[
\begin{aligned}
  P(N_4=2) 
  \quad = \quad & \, \|QUQUPUPU\psi\|^2 + \|PUQUQUPU\psi\|^2 
  \\
   + & \, \|PUPUQUQU\psi\|^2 + \|QUPUQUPU\psi\|^2 
  \\
   + & \, \|PUQUPUQU\psi\|^2 + \|QUPUPUQU\psi\|^2.
\end{aligned}
\]
The initial state in all our examples will be
\[ 
 \psi = \frac1{\sqrt2}\bigg(|0\>\otimes\uk + \imath \, |0\>\otimes\dk\bigg), 
\]
where $\iota$ is the imaginary unit.

\medskip

More generally, the probability of finding the walker $r$ times on the
positive subspace in the process of taking $n$ steps is given by the sum
\[
  P(N_n=r)\quad = \kern-3pt
  \sum_{\substack{{P_i=P\,\text{or}\,Q} \\ P_i=P\text{ for } r \text{ indices}}} 
  \kern-3pt \|(P_nU)\cdots(P_2U)(P_1U)\psi\|^2.
\]
These quantities will be plotted against the relative number of steps
$r/n$ in the graphics corresponding to the examples of the next
sections. We will represent both, the ``density" $P(N_n=r)$ and the
``cumulative distribution" $P(N_n\le r)=\sum_{k=0}^rP(N_n=k)$, which
should be compared with Figure~\ref{fig:clas}. Since we will be
dealing with values of $n$ in the range $90$--$8100$ and $r$ between
$0$ and $n$, it is clear that one needs a smart way to evaluate the
sum given above---a direct approach would involve a prohibitive number
of terms. Details on reducing the complexity of this calculation from
$O(2^n)$ to $O(n^3)$ will be given in \cite{GVW}.

\medskip 

The walker will evolve with a CMV walk with a doubly infinite CMV
matrix that treats right and left ``just the same", to have a ``fair"
comparison with the classical case of a fair coin. Restricting for
simplicity to real valued Verblunsky coefficients $\alpha_i$, we need
to impose
\begin{equation} \label{eq:fair}
  \alpha_i= \alpha_{-2-i}, \qquad i=0,1,2,3,\dots
\end{equation} 
a condition that leaves $\alpha_{-1}$ free. In the case of coined
walks this amounts to taking real coins $C_i$ such that
$C_i=C_{-1-i}$.

\medskip

Our results, displayed in the next few sections, illustrate the fact
that the quantum case gives us an embarrassment of riches, i.e. widely
different behaviours can take place. All of them are unique to the
quantum world, and yet some of them mimic the behaviour that most
people expect in the classical case.

\section{A look at the Hadamard walk}

The first time that the occupation problem was considered for the
Hadamard walk is \cite{Ko1}. This is a coined walk with constant coin
\begin{equation} \label{eq:Had}
  \frac{1}{\sqrt{2}} \begin{pmatrix} 1 & 1 \\ 1 & -1  \end{pmatrix}.
\end{equation}
Although this coin is not of the form \eqref{eq:coin}, such a form may
be obtained after a change of phases in the basis \cite{CGMV,CGMV2}.

Our approach differs from the one in \cite{Ko1}, which needs an ad hoc
normalization to get true probabilities due to the lack of a
connection with a real measurement protocol. Therefore, for a finite
number of steps our results do not agree. Nevertheless, we give some
evidence that the limiting results apparently---and
surprisingly---agree. Clarifying whether this coincidence holds by
chance or there are good reasons for it is something that deserves
future research.

\begin{figure}[t]
  \centering
  \includegraphics[width=.95\linewidth]{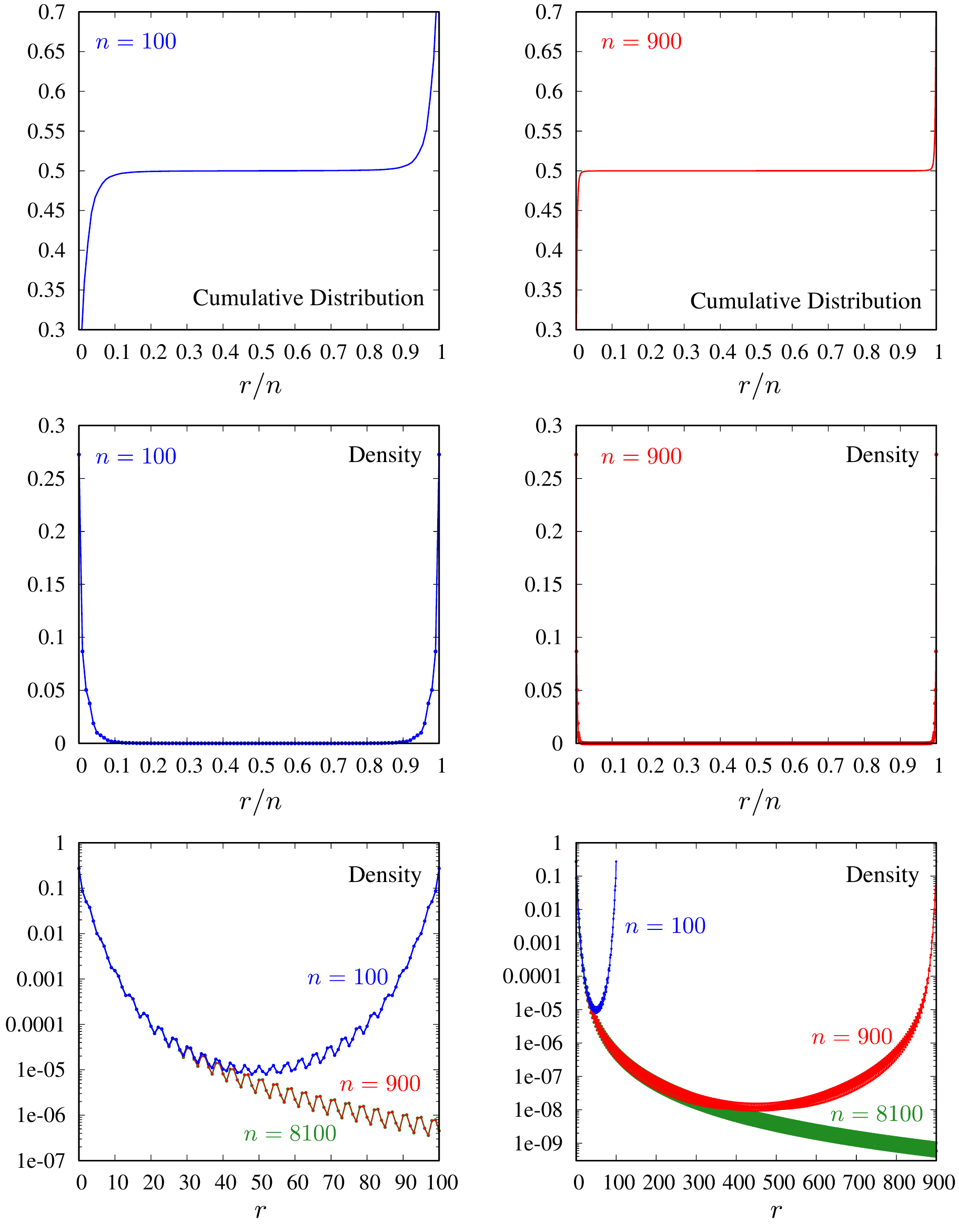}
  \caption{Cumulative distribution and density for $100$, $900$
    and $8100$ steps
    of the Hadamard walk (monitoring approach).}
  \label{fig:Had} 
\end{figure}

The monitoring approach to occupation times in the Hadamard walk leads
to the results shown in Figure~\ref{fig:Had}, which plots the density
and cumulative distributions for $100$, $900$ and $8100$ time steps.
This figure points to a limiting distribution function given by two
symmetric deltas at the edges of the interval. The take
home message is that, in the limit of an infinite number of steps, the
walker is always in the positive subspace or always in the negative
subspace, each with probability $1/2$. More precisely, the discrete
probabilities $P(N_n=r)$ appear to have a limit as $n\rightarrow\infty$
with $r$ held fixed,
\begin{equation}
  P_\infty(r)=\lim_{n\rightarrow\infty}P(N_n=r) =
  \lim_{n\rightarrow\infty}P(N_n=n-r).
\end{equation}
As seen in the bottom two panels of Figure~\ref{fig:Had}, this limit
has already nearly been reached for $0\le r\le 100$ by the time
$n=900$. Indeed the $n=900$ result lies directly on top of the
$n=8100$ result for $0\le r\le 100$. The numerical value of
$P(N_n\le100 \text{ or }N_n\ge n-100)$ with $n=900$ is $0.99994613$
while that of $n=8100$ is $0.99994530$. The reported digits with
$n=8100$ appear to have stabilized, i.e.~increasing $n$ further does
not change the first eight digits of the probability. So if $n$ is a
billion, the probability that the walker will be found on one side all
but at most 100 times is $0.99994530$, and the fraction of the time it
spends on the other side is $\le 100/10^9$, effectively zero.

\begin{figure}[t]
    \centering
    \includegraphics[width=.95\linewidth]{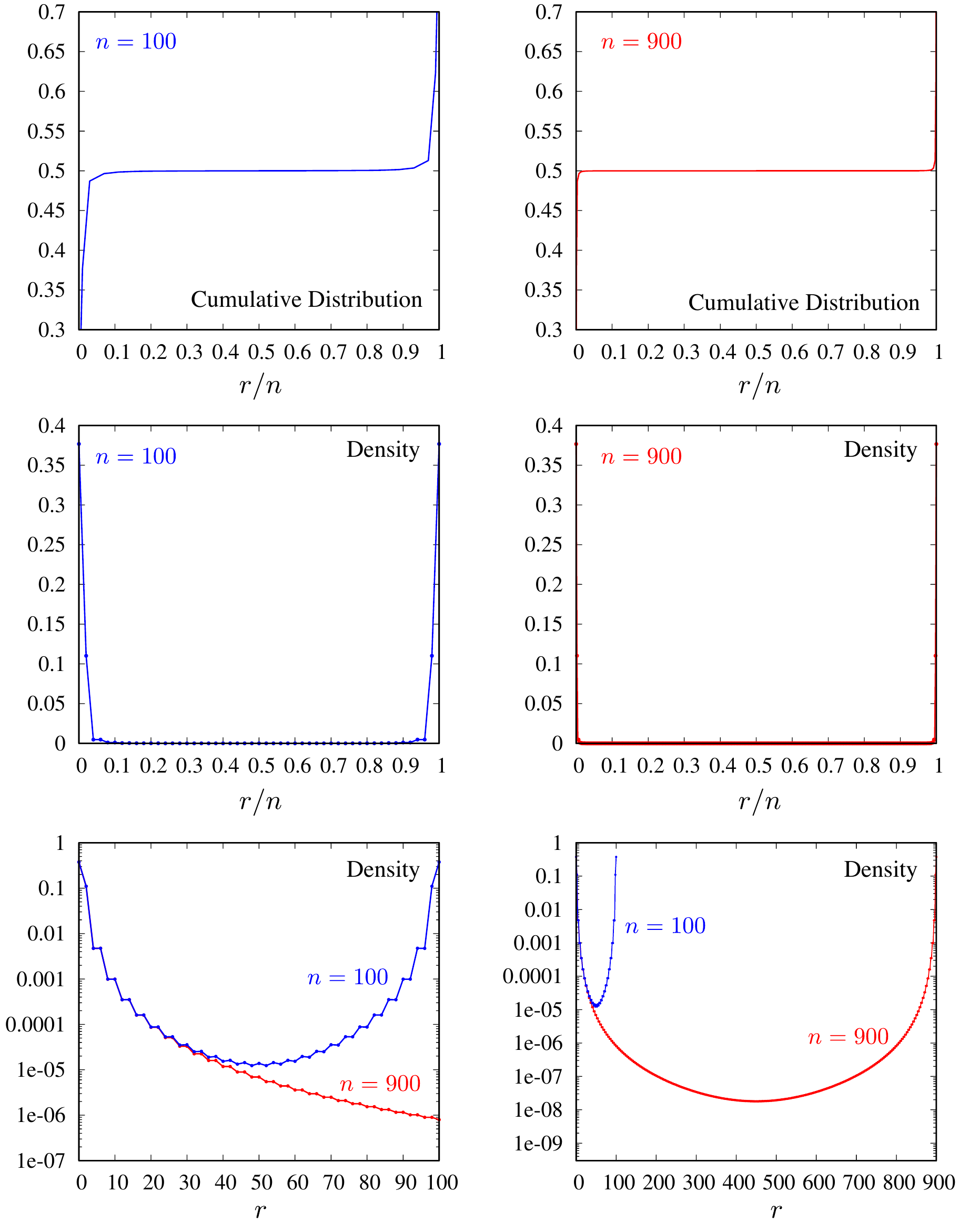}
    \caption{Cumulative distribution and density for $100$ and
      $900$ steps of the Hadamard walk (following \cite{Ko1}).}
    \label{fig:Had-K} 
\end{figure}

Using the alternative definitions and approach in \cite{Ko1} we have
computed the density and the corresponding cumulative distribution for
$n=100$ and $n=900$ steps of the Hadamard walk. The computation was
too expensive to proceed past $n=900$. The setup is different, and
only even values of $r$ lead to positive probabilities. Plots of
the cumulative distribution and density (i.e.~discrete probabilities)
are given in Figure~\ref{fig:Had-K}. This figure suggests the same
limit distribution and behavior as the monitoring approach, though
the values of $P_\infty(r)$ will be different.

\section{The walk with a constant coin}

In this section we deal with coined walks with a constant coin 
\begin{equation} \label{eq:c-coin}
  C = \begin{pmatrix} \rho & -\alpha \\ \alpha & \;\rho \end{pmatrix},
  \qquad \alpha\in(-1,1), \qquad \rho=\sqrt{1-|\alpha|^2}.
\end{equation}
Here we consider four different cases, corresponding to four choices
of $\alpha$, namely $3/5$, $12/13$, $143/145$ and $399/401$. In each
case the number of times steps is $120$. The plots for the occupation
times using the monitoring approach are given in
figures~\ref{fig:3/5} and \ref{fig:143/145} below.

\begin{figure}[h]
    \centering
    \includegraphics[width=\linewidth]{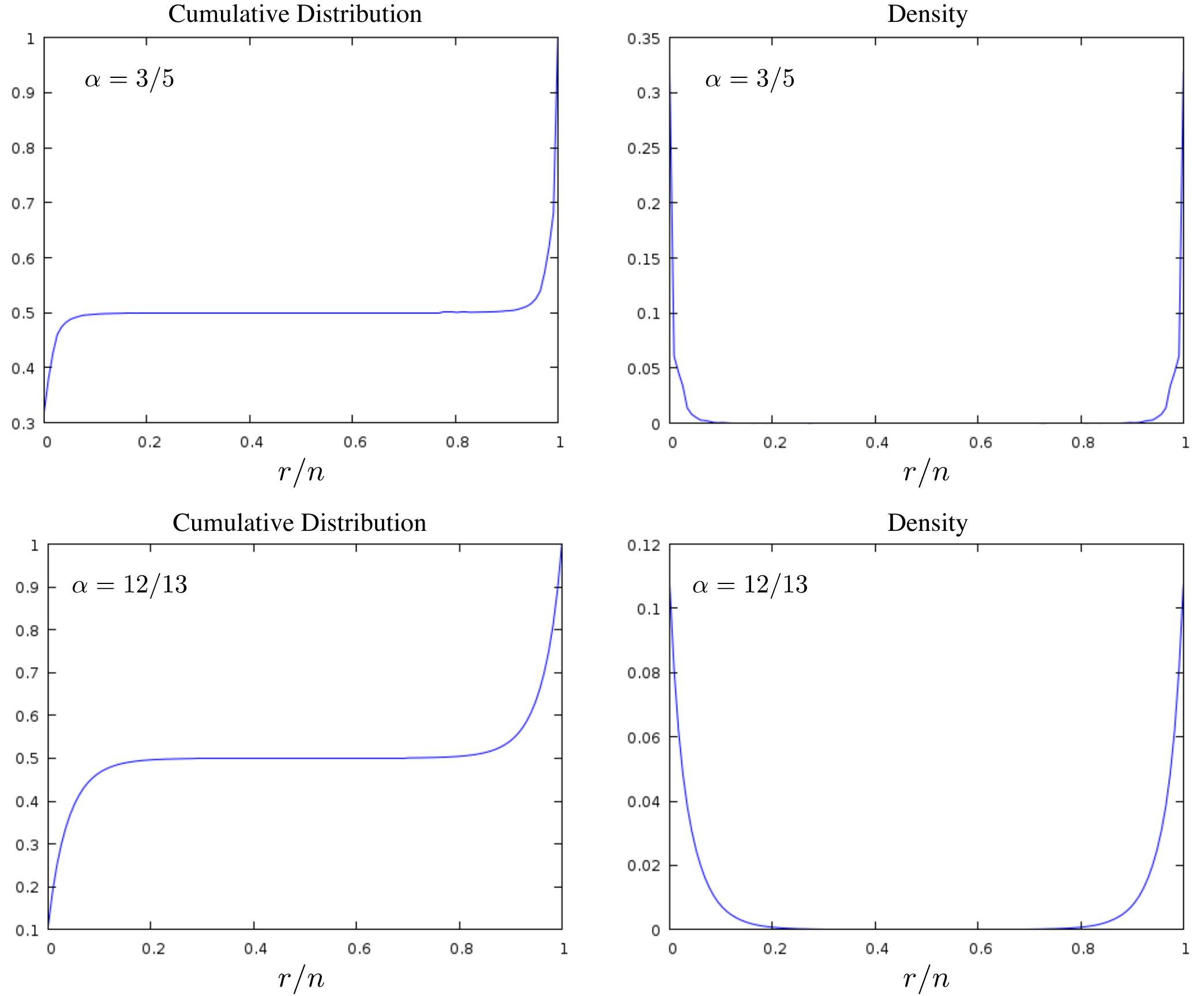}
    \caption{Cumulative distribution and density for the coin
      \eqref{eq:c-coin} with $\alpha=3/5$ and $\alpha=12/13$,
      both with 120 steps.}
    \label{fig:3/5} 
\end{figure}

\begin{figure}[h]
    \centering
    \includegraphics[width=\linewidth]{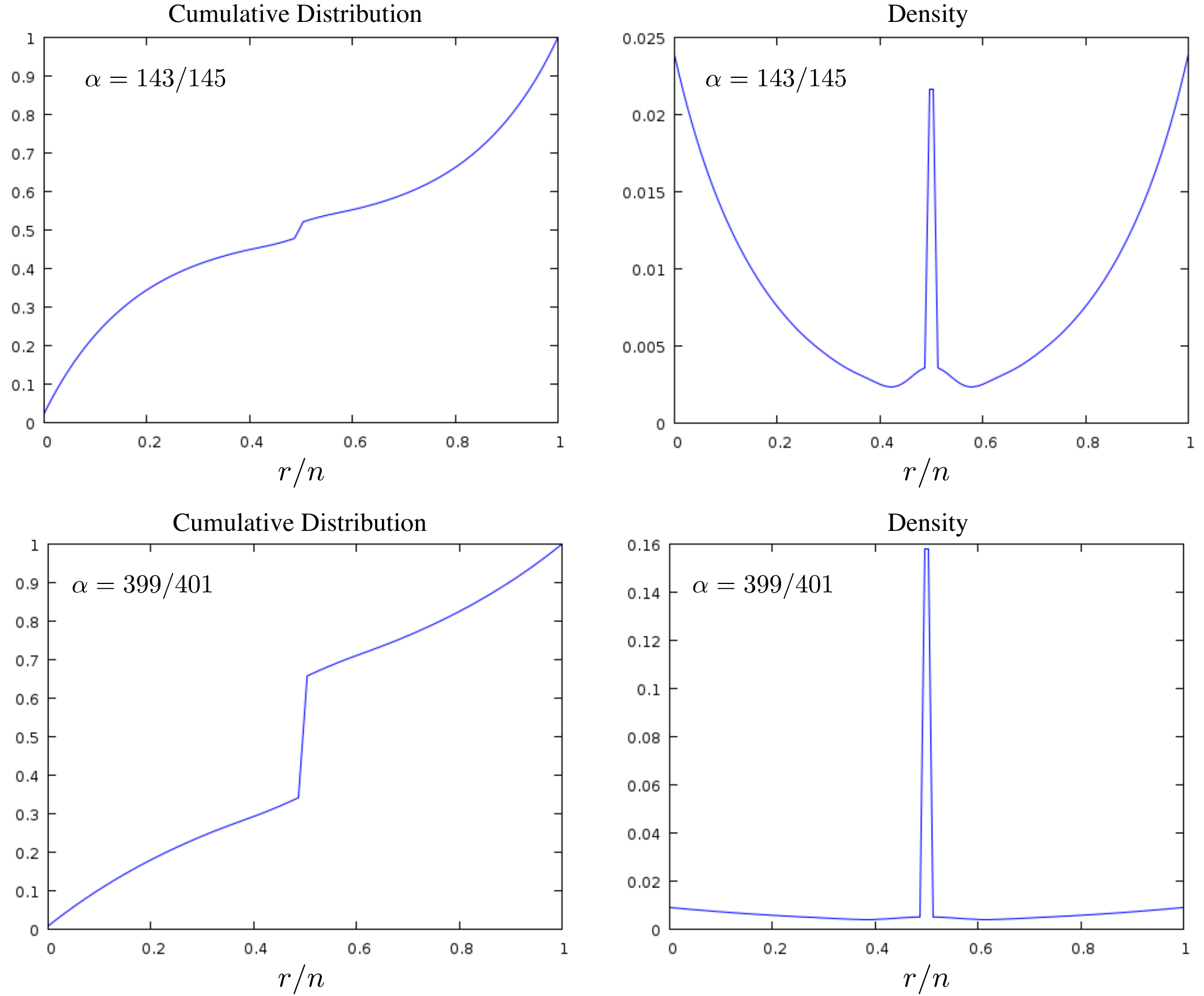}
    \caption{Cumulative distribution and density for the coin
        \eqref{eq:c-coin} with $\alpha=143/145$ and $\alpha=399/401$,
        both with 120 steps.}
      \label{fig:143/145} 
\end{figure}

The message here is that for constant coins that are different from
the Hadamard one the behaviour can differ substantially from that of
the Hadamard case.

\section{The even Verblunsky coefficients tend to one}

In this section we look at a CMV walk whose Verblunsky coefficients
$\alpha_{2i}\in(-1,1)$ depend on the site $i$, approaching the value
$1$ as $|i|$ tends to infinity. For simplicity we take the odd
Verblunsky coefficients to vanish, so as to have a coined walk with
site dependent coins $C_i$ of the form \eqref{eq:coin} such that
\begin{equation} \label{eq:coin-conv}
 C_i \quad \xrightarrow{|i|\to\infty} \quad
 C_\infty = \begin{pmatrix} 0 & -1 \\ 1 & \; 0 \end{pmatrix}.
\end{equation}

\begin{figure}[t]
    \centering
    \includegraphics[width=\linewidth]{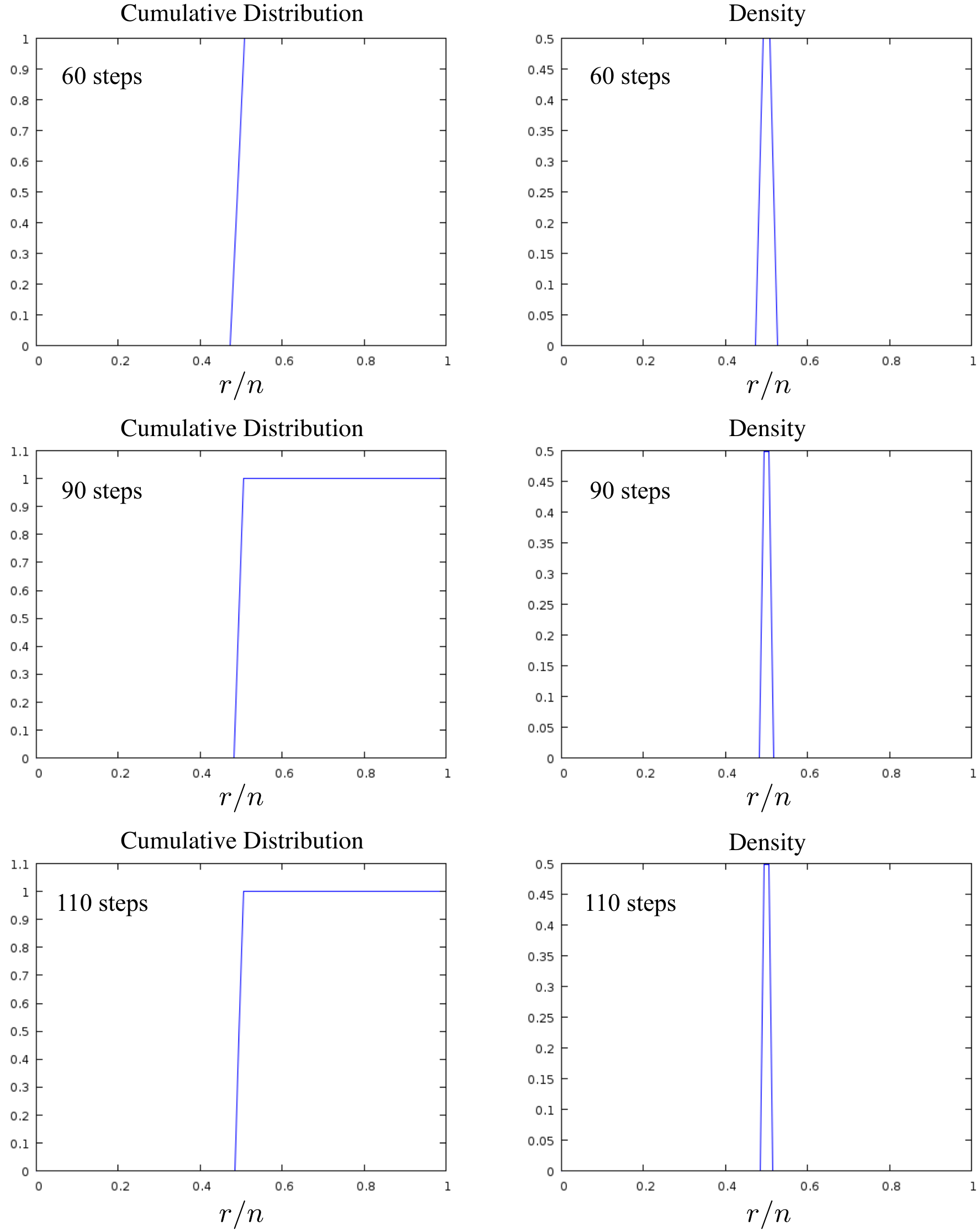}
    \caption{Cumulative distribution and density for $60$, $90$
      and $110$ steps.}
    \label{fig:60} 
\end{figure}

The three different plots we show differ in the number of time
steps. The plots are given in figure~\ref{fig:60}, and the effect of
increasing the number of steps is to make the density more and more
sharply concentrated around the value $1/2$.

The take home message is that in the limit of an infinite number of
steps, with probability $1$ the walker spends exactly half of the the
time on the positive subspace and half on the negative subspace. In
spirit, this is what most people feel the classical coin should be
doing.

\medskip
 
In this example, the Verblunsky coefficients are chosen to satisfy 
\[
  \alpha_{2i-1}=0,
  \qquad\qquad
  \alpha_{2i}\; \xrightarrow{|i|\to\infty} \; 1. 
\]
More explicitly, if $i$ is non-negative and even then
\begin{equation*}
  \alpha_i = \frac{(i+1)^{10}-1}{(i+1)^{10}+1}, \qquad\qquad
  \rho_i = \frac{2(i+1)^5}{(i+1)^{10}+1},
\end{equation*}
and if $i$ is odd then $\alpha_i=0$ and $\rho_i=1$.

The corresponding CMV matrix $U$ is therefore a compact perturbation
of the unitary matrix obtained by setting $\alpha_{2i-1}=0$ and
$\alpha_{2i}=1$ in a CMV, which turns out to be a direct sum of the
blocks $C_\infty$ in \eqref{eq:coin-conv}. As a consequence of Weyl's
theorem on the essential spectrum, the spectrum of $U$ accumulates on
the eigenvalues $\pm i$ of $C_\infty$, so that it is pure point.

A couple of natural questions arise: Is the naive behaviour of
occupation times observed in this example a common feature of any CMV
walk with pure point spectrum? How does the occupation time
distribution change when considering a CMV walk with other kind of
singular spectrum?

While the first question remains as a challenge, in the next section
we explore the second one.

\section{A look at the Riesz walk}

In this section we deal with the singular continuous measure
constructed by F. Riesz \cite{Ri}, back in 1918. In \cite{GV-Riesz} we
study a purely singular continuous quantum walk in the non-negative
integers naturally associated to it, by considering the semi-infinite
CMV matrix generated by the Riesz measure. In this paper we consider
an extension of this walk to the integers.

The measure on the unit circle that F. Riesz built is formally given by the expression
\begin{equation} \label{RIESZ}
  d\mu(z) = \prod_{k=1}^{\infty} (1+ \cos(4^{k} \theta))
  \frac{d\theta}{2\pi}
  = \prod_{k=1}^{\infty} (1+(z^{4^{k}}+z^{-4^{k}})/2) \frac{dz}{2\pi iz}.
\end{equation}
Here $z=e^{i \theta}$. If one truncates this infinite product the
corresponding measure has a nice density. These approximations
converge weakly to the Riesz measure.

The construction of this example is based on detailed knowledge of the
coefficients $\alpha_i$, $i\ge0$, for the Riesz measure which are
found in \cite{GV-Riesz}. We define a ``Riesz walk" on the integers as
a a CMV walk defined by extending the Verblunsky coefficients of the
Riesz measure to negative indices according to the ``fair" rule
\eqref{eq:fair}.

This example is of interest since it is not clear that a ``limit law"
exists for the ``site distribution" of the Riesz walk on the positive
half of the integers, see \cite{GV-Riesz}. The same issue arises when
one looks at the walk on the integers. The existence of such a limit
law for the case of a constant coin is discussed in \cite{Ko}.

\begin{figure}[h]
    \centering
    \includegraphics[width=\linewidth]{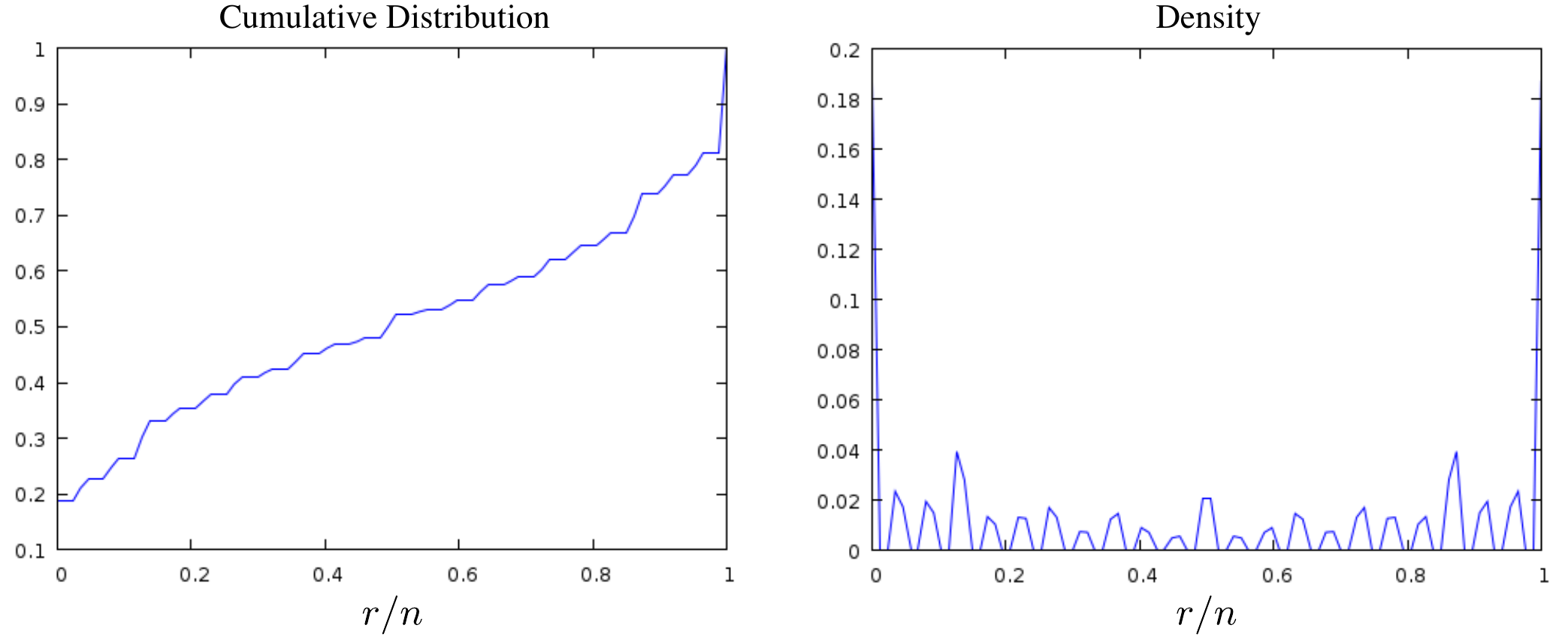}
    \caption{Cumulative distribution and density for 90 steps of
      Riesz's walk.}
    \label{fig:Riesz90} 
\end{figure}

Figure~\ref{fig:Riesz90} shows the density and cumulative
distribution for 90 steps of the Riesz walk on the integers.

These examples open up many interesting questions: What can be said in
general about the occupation time distribution in CMV walks? Is there
any spectral characterization for the different behaviours of such
distributions? Is the connection with orthogonal polynomials on the
unit circle useful for such a characterization? In particular, is
there any important role reserved for the Schur functions in this
respect? The last question is motivated by the central role played by
Schur functions in both, the theory of orthogonal polynomials on the
unit circle \cite{Si04-1,Si04-2}, and the study of quantum walks
\cite{GVWW,bourg,CGMV2,GV-Schur,GLV,toph}.







\printindex
\end{document}